\documentclass[aps,preprint,prd,floatfix,nofootinbib]{revtex4-1}%
\usepackage{amsfonts}
\usepackage{amsmath}
\usepackage[utf8]{inputenc}
\usepackage{amssymb}
\usepackage{float}
\usepackage{xcolor}
\usepackage{graphicx}%
\providecommand{\U}[1]{\protect\rule{.1in}{.1in}}

\begin{document}
\title{Solitons in a cavity for the Einstein-$SU(2)$ Non-linear Sigma Model and Skyrme model}
\author{Alex Giacomini$^{1}$, Marcela Lagos$^{2}$, Julio Oliva$^{2}$, Aldo Vera$^{2}$}
\address{
$^1$ Instituto de Ciencias Físicas y Matemáticas, Universidad Austral de Chile, Valdivia, Chile.\\
$^2$ Departamento de Física, Universidad de Concepción, Casilla, 160-C, Concepción, Chile.}
\email{alexgiacomini@uach.cl, marcelagos@udec.cl, juoliva@udec.cl, aldovera@udec.cl}
\begin{abstract}
In this work, taking advantage of the Generalized Hedgehog Ansatz, we construct new self-gravitating solitons in a cavity with mirror-like boundary conditions for the $SU(2)$ Non-linear Sigma Model and Skyrme model. For spherically symmetric
spacetimes, we are able to reduce the system to three independent equations
that are numerically integrated. There are two branches of well-behaved solutions. The first branch is defined for arbitrary values of the Skyrme coupling and therefore also leads to a gravitating soliton in the Non-linear Sigma Model, while the second branch exists only for non-vanishing Skyrme coupling. The solutions are quasi-static and in the first branch are characterized by two integration constants that correspond to the frequency of the phase of the Skyrme field and the value of the Skyrme profile at the origin, while in the second branch the latter is the unique parameter characterizing the solutions. These parameters determine the size of the cavity, 
the redshift at the boundary of the cavity, the energy of the scalar field and the charge associated to a $U(1)$ global symmetry. We also show that within this ansatz, assuming analyticity of the matter fields, there are no spherically symmetric black hole solutions.

\end{abstract}
\maketitle

\section{Introduction}
Non-linear Sigma models appear in many contexts, as for example to describe the dynamics of Goldstone bosons \cite{Nair:2005iw}, in condensed matter systems \cite{Manton:2004tk}, in supergravity \cite{Freedman:2012zz},
as well as being the building blocks of classical string theory. In the case of light mesons, it can be shown that the low energy dynamics can be correctly described by a Non-linear Sigma Model for $SU(2)$. In such low energy 
processes, the mesons can be seen as Goldstone bosons. In flat spacetime, the inclusion of the Skyrme term allows to construct static regular solitons with finite energy, which describe baryons \cite{Adkins:1983ya}. 
In the latter scenario the ansatz for the $SU(2)$ group element is given by $U_{\text{sol}}=\exp({iF(r)\vec{\tau}\cdot\hat{x}})$, with $\vec{\tau}$ the $SU(2)$ generators. A more general ansatz is defined by the Generalized 
Hedgehog Ansatz, which includes $U_{\text{sol}}$ as a particular case, and is defined by 
\begin{equation}
U^{\pm 1}=Y^0\mathbf{1}\pm Y^{i}t_{i}\ ,
\end{equation}
where $\mathbf{1}$ is the $2\times 2$ identity matrix and
\begin{equation}
Y^{0}=\cos\alpha(x^\mu)\ ,\ \ Y^{i}=\widehat{n}^{i}\sin\alpha(x^\mu) \ , \qquad(Y^{0}%
)^{2}+Y_{i}Y^{i}=1 \ , \label{general-Y}%
\end{equation}
with a generalized radial unit vector
\begin{equation}
\widehat{n}^{1}=\cos \Theta(x^\mu)\sin F(x^\mu)\ ,\ \ \ \widehat{n}^{2}=\sin \Theta(x^\mu) \sin F(x^\mu)\ ,\ \ \ \widehat
{n}^{3}=\cos F(x^\mu) \ . \label{general-Y2}%
\end{equation}
Here $\alpha, \Theta$ and $F$ are arbitrary functions of the space-time coordinates. This ansatz was originally introduced in the context of the Gribov problem in regions with non-trivial 
topology \cite{Canfora:2013mrh}, and has been shown to provide a very fruitful arena to construct new solutions of the theory. In reference \cite{canforamaeda}, the compatibility of this ansatz on the Einstein-Skyrme 
theory was thoroughly explored considering a space-time which is a warped product of a two-dimensional space-time with an Euclidean constant curvature manifold. Also, within this ansatz, a novel non-linear superposition 
law was found in \cite{Canfora:2013xja} for the Skyrme theory, which was latter extended to the curved geometry of AdS$_2\times S^2$ in reference \cite{Tall}. Even more, the ansatz allows for exact solitons with a kink 
profile \cite{Chen:2013qha}. Asymptotically AdS wormholes and bouncing cosmologies with self-gravitating Skyrmions were constructed in \cite{Ayon-Beato:2015eca} as well as other time dependent cosmological solutions 
with non-vanishing topological charge \cite{Paliathanasis:2017cie}. Also within the context of the generalized hedgehog ansatz, for the $SU(2)$ Non-linear Sigma Model, topologically non-trivial gravitating solutions were constructed 
in \cite{Canfora:2017gno} which cannot decay on the trivial vacuum due to topological obstructions and, more recently, planar asymptotically AdS hairy black hole solutions were found in \cite{Astorino:2017rde}.
\\
\\
In this paper we will explore a new family of solutions within the Generalized Hedgehog Ansatz which describe spherically symmetric, quasi-static configurations in a cavity. By imposing mirror-like boundary 
condition for the matter field we numerically construct new self-gravitating solitons for the $SU(2)$ Skyrme model and Non-linear Sigma Model. In Section II we introduce the Generalized Hedgehog Ansatz.
In Section III we reduce the system to three non-linear equations and argue that in order to have configurations with finite energy it is necessary to introduce a mirror at a finite proper distance 
from the origin. Section IV is devoted to the numerical integration of the system that leads to two well-behaved branches. The first branch is well behaved for arbitrary values of the coupling constant of 
the Skyrme term $\lambda$, while the second leads to well-behaved solutions only for non-vanishing $\lambda$. Section V contain the conclusions and further comments as well as the proof that, within this ansatz, 
there are no 
black holes supported by an analytic Skyrme field.
\section{The $SU(2)$ Einstein-Skyrme and Einstein-Nonlinear sigma model}
In this paper we will be concerned with the gravitating Einstein-Skyrme model as well as with the Einstein-Non-linear Sigma Model systems. The action is given by
\begin{align}
I[g,U]  &  = \int d^{4}x \sqrt{-g}\left(  \frac{R}{2\kappa
}+\frac{K}{4} \text{Tr}\left(A^{\mu}A_{\mu}+\frac{\lambda}{8} F_{\mu\nu}F^{\mu\nu}\right) \right)  ,
\end{align}
where $R$ is the Ricci scalar, $A_{\mu}:=U^{-1}\nabla_{\mu}U$ and $F_{\mu\nu}=[A_\mu,A_\nu]$.  Here $U$ is a scalar field valued in $SU(2)$ and therefore $A_{\mu}=A_{\mu}^{i} t_{i}$, 
with $t_{i}=-i\sigma_{i}$ the $SU(2)$ generators, $\sigma_{i}$ being the Pauli matrices. We work in the mostly plus signature, Greek
and Latin indices run over spacetime and the algebra, respectively. Hereafter without loosing generality we set $K=1$.

The field equations for this theory are the Einstein equations
\begin{align}
\mathcal{E}_{\mu\nu} =G_{\mu\nu}-T_{\mu\nu}=0\ , \label{eeqsfirst}
\end{align}
with energy-momentum tensor given by
\begin{align*}
T_{\mu\nu}  &  =-\frac{1}{2}\text{Tr}\left(A_{\mu}A_{\nu}-\frac{1}{2}%
g_{\mu\nu}A_{\alpha}A^{\alpha}+\frac{\lambda}{4}(g^{\alpha\beta}F_{\mu\alpha}F_{\nu\beta}-\frac{1}{4}g_{\mu\nu}F_{\alpha\beta}F^{\alpha\beta})\right)  ,
\end{align*}
satisfying the dominant energy condition \cite{Gibbons:2003cp}, and the Skyrme equations
\begin{align}
\nabla^{\mu}A_{\mu}+\frac{\lambda}{4}\nabla^\mu[A^\nu,F_{\mu\nu}]=0.
\label{nlsmeqs}
\end{align}
We will consider the generalized hedgehog ansatz (\ref{general-Y}) and (\ref{general-Y2}) with $F\left(x^\mu\right)=\frac{\pi}{2}$. 
The functions $\alpha$ and $\Theta$ of the ansatz (\ref{general-Y}) and (\ref{general-Y2}) are scalar functions: $\alpha$ describes the energy profile of the configuration while $\Theta$\ describes its orientation in isospin space. One can check that the
above ansatz has vanishing baryon charge, thus we are within the pionic sector. The group manifold of $SU(2)$ is the three-sphere $S^3$, and our ansatz turns on the field along the $S^2\subset S^3$ submanifold. The advantage of the Generalized Hedgehog Ansatz is given by the fact that the Skyrme equations reduce to a single equation provided \cite{canforamaeda},
\begin{eqnarray}
\square\Theta   =0\  ,\nabla_{\mu}\Theta\nabla^{\mu}\alpha   =0\ ,\label{constraints1}\\
(\nabla^{\mu}\nabla^{\nu}\Theta)\nabla_\mu\Theta\nabla_\nu\Theta=0\ ,\\
(\nabla^{\mu}\nabla^{\nu}\alpha)\nabla_\mu\alpha\nabla_\nu\Theta=0\ .
\label{constraints3}
\end{eqnarray}
Even though these equations may seem too restrictive, we will show below that they are compatible with the existence of quasi-static solitonic solutions in a cavity.

With this, the Einstein and Skyrme equations reduce to
\begin{equation}\label{eeqs}
\mathcal{E}_{\mu\nu}=G_{\mu\nu}-\kappa T_{\mu\nu}=0\ ,
\end{equation}
with
\begin{align}
T_{\mu \nu }=&\biggl[(\nabla _{\mu }\alpha )(\nabla _{\nu }\alpha )+\sin
^{2}\alpha (\nabla _{\mu }\Theta )(\nabla _{\nu }\Theta )  +\lambda \sin ^{2}\alpha \notag \\
&\times  \biggl((\nabla \Theta )^{2}(\nabla _{\mu }\alpha
)(\nabla _{\nu }\alpha )+(\nabla \alpha )^{2}(\nabla _{\mu }\Theta )(\nabla
_{\nu }\Theta )\biggl)  \notag \\
& -\frac{1}{2}g_{\mu \nu }\biggl((\nabla \alpha )^{2}+\sin ^{2}\alpha
(\nabla \Theta )^{2}+\lambda \sin ^{2}\alpha (\nabla \Theta )^{2}(\nabla
\alpha )^{2}\biggl)\biggl]\ ,
\end{align}
supplemented by
\begin{align}
&\square\alpha-\frac{1}{2}\sin (2\alpha )(\nabla \Theta )^{2} 
+\lambda \biggl[(\nabla _{\mu }\alpha )\nabla ^{\mu }\biggl(\sin
^{2}\alpha (\nabla \Theta )^{2}\biggl) \notag \\
&+\sin ^{2}\alpha (\nabla \Theta
)^{2}(\square\alpha )-\frac{1}{2}\sin (2\alpha )(\nabla \alpha
)^{2}(\nabla \Theta )^{2}\biggl]=0\ .  \label{skyrme-eq}
\end{align}
These equations can also be obtained from the effective action
\begin{equation}
I_{\text{eff}}=\int\sqrt{-g}\left[\frac{R}{2\kappa}-2\left(\partial_\rho\alpha\partial^\rho\alpha+\sin^2(\alpha) \partial_\rho\Theta\partial^\rho\Theta+\frac{\lambda}{2}\sin^2(\alpha)(\nabla\alpha)^2(\nabla\Theta)^2\right)\right], \label{effaction}\ 
\end{equation}
provided the constraints (\ref{constraints1})-(\ref{constraints3}) are fulfilled.
Einstein equations (\ref{eeqs}) and the Skyrme equation (\ref{skyrme-eq}) are obtained from the variation of $I_{\text{eff}}$ w.r.t. the metric and the scalar $\alpha$, respectively, and the equation for $\Theta$ is trivially satisfied after imposing the constraints (\ref{constraints1})-(\ref{constraints3}). The effective action, as well as the constraints, are invariant under the global transformation
\begin{align}
\delta_{\left(  1\right)  }\alpha & =0 \ , \qquad\delta_{\left(  1\right)
}\Theta=\epsilon \ ,\label{sim1}
\end{align}
where $\epsilon$ is a parameter. The symmetry transformation $\delta_{(1)}$ allows to construct a locally conserved current which, when integrated within the cavity, leads to a finite conserved charge.
\section{The system and its finite energy solutions}

We consider a static spherically symmetric space-time metric
\begin{equation}
ds^{2}=-f(r)dt^{2}+\frac{dr^2}{h(r)}+r^{2}\left(d\theta^{2} +\sin^{2}{\theta
}d\phi^{2}\right)\ , \label{metric}
\end{equation}
and the following dependence for the matter fields
\begin{align}
\alpha &  =\alpha(r) \ , \qquad\Theta=\omega t \ , \label{fields}
\end{align}
where $\omega$ is a frequency, leading to a time independent energy-momentum tensor and therefore the whole configuration is quasi-static\footnote{The kinetic term for the Non-linear Sigma Model
, $(\partial\alpha)^2+\sin^2\alpha(\partial\Theta)^2$, is mapped to $\left(1+|\Phi|^2/4\right)^{-2}|\partial\Phi|^2$ with $\Phi=\rho\exp{(i\chi)}$ via the transformation $\Theta=\chi$ and $\alpha=\arccos\left(\frac{4-\rho^2}{4+\rho^2}\right)$. This makes explicit the fact that $\Theta$ is a phase that, according to our ansatz, rotates in time at a frequency $\omega$ with respect to the coordinate time $t$.}. For this ansatz, the constraint equations (\ref{constraints1})-(\ref{constraints3}) are automatically fulfilled and the Einstein-Skyrme system reduces to three independent equations. 
We work with the equations $\mathcal{E}_{tt}$ and $\mathcal{E}_{rr}$ (with $\mathcal{E}_{\mu\nu}$ defined in (\ref{eeqs})), as well as (\ref{skyrme-eq}). 
Introducing for simplicity $u(r)=\sin\alpha(r)$, and setting $2\kappa=1$ one obtains the following non-linear system
\begin{widetext}
\begin{align}
r^2\omega^2u^2\left( u^2-\lambda h u'^2 -1 \right) +f\left( 2(u^2-1)(rh'-1) +h(2u^2- r^2u'^2-2 \right) &= 0\ ,\label{EE1}  \\
 f\left( 2(1-u^2)+h\left( 2(u^2-1) +r^2u'^2\right)  \right) && \notag \\
 -r\left( r\omega^2u^4 +2hf'+u^2( r\lambda\omega^2 hu'^2 -2hf'- r\omega^2)   \right) &=0\ , \\
-2r\omega^2fu(u^2-1)^2+(u^2-1)\left( 2f^2(1+h)-4\lambda\omega^2fhu^2+r^2\lambda\omega^4u^4   \right)u' &&  \notag \\
+2r(\lambda\omega^2-f)fhuu'^2 -2rfh(u^2-1)\left( \lambda\omega^2u^2 -f\right)u'' &=0\ . \label{EE3}
\end{align}
\end{widetext}

It is worth pointing out that the parameter $\omega$ can be absorbed in the field equations by rescaling the radial coordinate as well as the Skyrme coupling in the form $r \rightarrow \bar{r}=\omega r,\ \lambda \rightarrow \bar{\lambda}=\omega^2 \lambda$. While in the Non-linear Sigma Model this transformation reduces the number of independent parameters to be provided before numerical integration, in the presence of the Skyrme term the freedom in $\omega$ is mapped to the freedom to choose the value of $\bar{\lambda}$.

Assuming that $\alpha$ goes to zero as $r$ goes to infinity, on an asymptotically flat space-time, equation (\ref{EE1}) reduces to
\begin{equation}
r\alpha''(r)+2\alpha'(r)+r\omega^2\alpha(r)=0\ .
\end{equation}
Consistently, this equation is equivalent to the equation for the radial profile of a massless scalar field in Minkowski space-time, and admits the following asymptotic behavior $\alpha(r)\rightarrow\cos(\omega r)/r$ as $r$ goes to infinity.

It is a straightforward computation to show that this asymptotic behavior is not compatible with having a finite mass. If we want to construct gravitating solitons in this sector of the Generalized Hedgehog Ansatz for the Skyrme model, it turns out to be necessary to confine the system into a cavity. In what follows we do so, by imposing mirror-like boundary conditions for the profile $\alpha$ at a finite value of the radial coordinate $r=r_m$, i.e. we impose the boundary condition $\alpha(r_m)=0$. As shown in \cite{Alvarez:2017cjm}, the introduction of a finite box also allows to construct examples of Skyrmions-anti-Skyrmions bound states, as well as time crystals.
\\
A similar situation occurs for the Einstein-Maxwell system coupled to a massless charged scalar (see e.g. appendix A of \cite{Ponglertsakul:2016wae}). The asymptotic behavior of the scalar field is not compatible 
with the requirement of asymptotic flatness and finite mass for solitons and black holes, and one is therefore forced to enclose the system into a cavity. This system has been particularly 
fruitful for the study of the non-linear evolution of the superradiant instability due to the electric charge of a scalar field including a mass 
term  \cite{Herdeiro:2013pia}-\cite{Sanchis-Gual:2015lje} as well as
self-interaction \cite{Sanchis-Gual:2016tcm}, leading to  the formation of hairy black holes \cite{Sanchis-Gual:2016ros}. The system in a cavity allows for the existence of solitons as well as black holes, and in the previous references there have been observed dynamical evolutions in both directions in different regimes \footnote{As shown in \cite{Bosch:2016vcp}, a negative cosmological 
constant provides a setup to naturally implement an effective cavity, such that the superradiant instability in the spherically symmetric charged case leads to a hairy black hole even for a 
massless scalar, obtaining results that are qualitatively similar to those of the system enclosed in a cavity.}.

\section{The new solitons}

The requirement of having a regular center at $r=0$ leads to two soliton branches. The first corresponds to a branch analytic in the Skyrme coupling, with
\begin{subequations}\label{bc1}
\renewcommand{\theequation}{\theparentequation.\arabic{equation}}
\begin{align} \label{expansion1}
f_1(r)  &= f_0+\frac{1}{3}u_0^2\omega^2r^2+\frac{u_0^2\omega^4 \left( 6 f_0^2( 2u_0^2 -1) +f_0 u_0^2(7-19u_0^2 )\lambda\omega^2 +6u_0^6\lambda^2\omega^4 \right)}{180 f_0(f_0-u_0^2\lambda\omega^2)^2}r^4 +\mathcal{O}\left(r^{6}\right) \ , \\
h_1(r)&= 1-\frac{u_0^2\omega^2}{6f_0}r^2+\frac{u_0^2\omega^4\left( f_0^2(2+u_0^2) -2f_0u_0^2(2+u_0^2)\lambda\omega^2 +3u_0^6\lambda^2\omega^4 \right)}{90f_0^2(f_0-u_0^2\lambda\omega^2)^2}r^4  +\mathcal{O}\left(r^{6}\right) \ ,  \\
u_1(r)&= u_0+\frac{u_0(u_0^2-1)\omega^2}{6(f_0-u_0^2\lambda\omega^2)}r^2-\frac{u_0(u_0^2-1)\omega^4\left( f_0^2(3-7u_0^2) +5f_0u_0^2(1+u_0^2)\lambda\omega^2 -6u_0^6\lambda^2\omega^2 \right)}{360f_0(f_0-u_0^2\lambda\omega^2)^3}r^4 +\mathcal{O}\left(r^{6}\right) \ . 
\end{align}
\end{subequations}
The second branch, non-analytic in $\lambda$, is given by
\begin{subequations}\label{bc2}
\renewcommand{\theequation}{\theparentequation.\arabic{equation}}
\begin{align} \label{expansion2}
f_2(r)  &= \lambda u_0^2\omega^2 +\frac{11u_0^2\omega^2}{30} r^2 +\frac{4u_0(20u_0^2-21)\omega^2 }{225\sqrt{5 \lambda(1-u_0^2)}}r^3 +\mathcal{O}\left(r^{4}\right)\ , \\
h_2(r)&= 1-\frac{7}{30\lambda}r^2 +\frac{4(6-5u_0^2)}{75\lambda^{\frac{3}{2}}u_0\sqrt{5(1-u_0^2)}}r^3 +\mathcal{O}\left(r^{4}\right) \ , \\
u_2(r)&= u_0-\frac{\sqrt{1-u_0^2}}{\sqrt{5\lambda}}r+\frac{3-10u_0^2}{150\lambda u_0}r^2+\frac{261-1890u_0^2+2125u_0^4}{38250\lambda^{\frac{3}{2}}u_0^2\sqrt{5(1-u_0^2)}} r^3  +\mathcal{O}\left(r^{4}\right) \ . 
\end{align}
\end{subequations}
The latter solution is intrinsic to the presence of the Skyrme term.

These two branches define the data at the origin which, after numerical integration, will determine the data at the mirror located at $r=r_m$.

Note that for the first branch, for a given value of the Skyrme coupling, the free parameters are $f_0$, $u_0$ and $\omega$. Normalizing the time coordinate $t$ to coincide with the proper time of a geodesic observer located at the origin sets $f_0=1$. This region is shared by all the configurations and one can therefore compare their physical parameters in a consistent manner. We therefore fix $f_0=1$ in this branch. For the second branch, the value of the $-g_{tt}$ component of the metric at the origin is not a free parameter any more and is fixed by $f_2(0)=\lambda u_0^2\omega^2$. We can still normalize the time coordinate to coincide with the proper time of a geodesic observer located at the origin by introducing the scaling $t\rightarrow  t/\sqrt{\lambda u_0^2\omega^2}$. In this manner, the parameter $\omega$ is absorbed from all the functions and the rotation of the phase is locked in terms of the Skyrme coupling and the value of the scalar at the origin as $\Theta=\omega t \rightarrow \Theta=t/\sqrt{\lambda u_0^2}$. Equivalently, this is accomplished if we directly set $\omega=1/\sqrt{\lambda u_0^2}$ for the integration of this branch. In this manner, for a given value of the Skyrme coupling, $u_0$ is the unique parameter characterizing the solutions in the second branch.

For the numerical integration we proceed as follows: We fix the coordinate $t$ to be the proper time of an observer at the origin, which leaves us with two ($\omega$ and $u_0$) and one ($u_0$) free parameter for the first and second branch, respectively. Then, for both branches, we integrate the system (\ref{EE1})-(\ref{EE3}) from a regulator $\epsilon\sim 0$ outwards, using the initial conditions for radial integration that 
come from expansion (\ref{bc1}) and (\ref{bc2}) up to order $\mathcal{O}(r^8)$. We locate the radius of the mirror at the first zero of the Skyrme field $u(r)$. Finally, the functions obtained after the integration are used to compute the energy and the $U(1)$ charge which are respectively given by
\begin{align}
M=-4\pi\int_{0}^{r_m}T^t_{\ t}r^2 dr=8\pi r_m(1-h(r_m)) \ ,
\label{masa}%
\end{align}
and
\begin{equation}
Q=-16\pi \omega \int_{0}^{r_m} dr \frac{ r^2}{\sqrt{f(r)h(r)}}\left(1+\frac{\lambda}{2}\alpha'^2 h(r)\right)\sin(\alpha)^2\ . \label{charge}
\end{equation}
Below, we present the results of the integration for each branch.

\subsection{Branch 1: Analytic in $\lambda$}
\begin{figure}[H]
\centering
\includegraphics[scale=0.55]{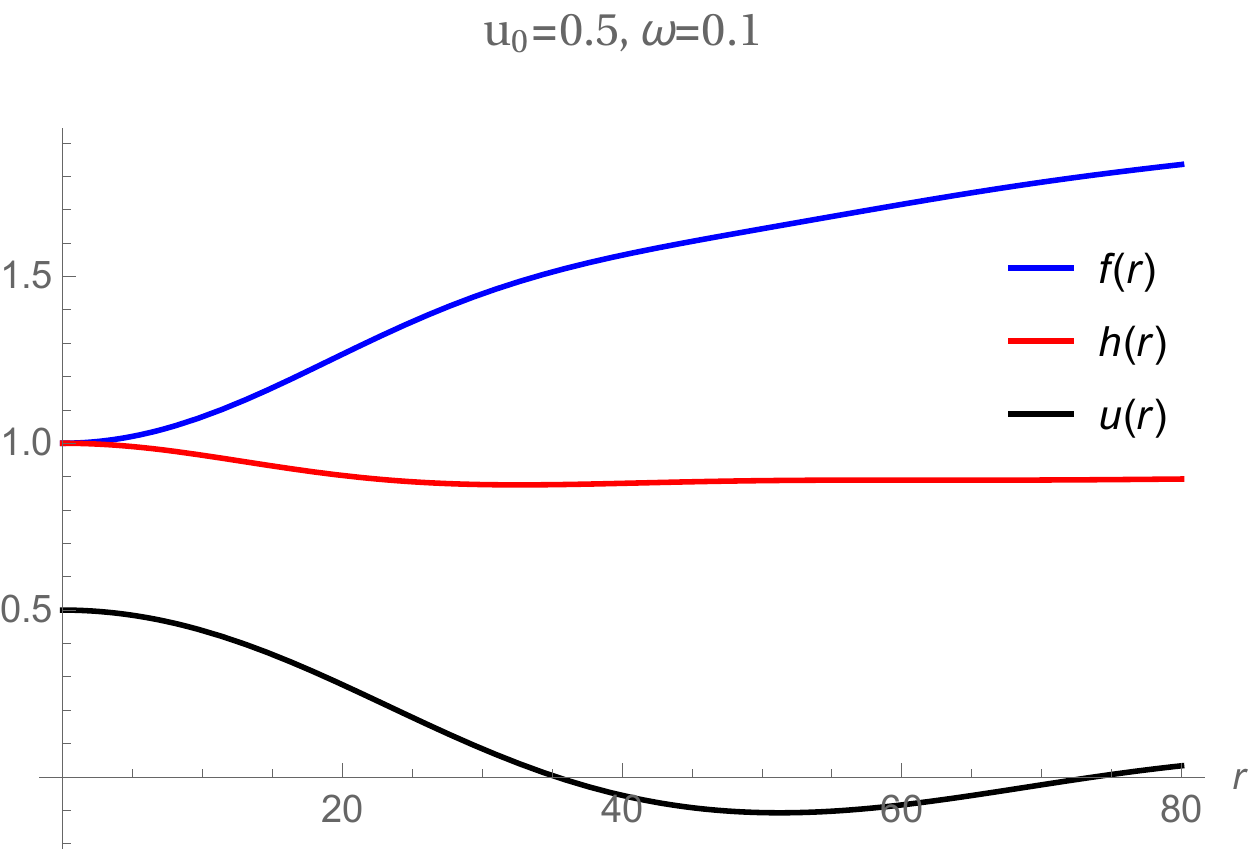}\quad\includegraphics[scale=0.55]{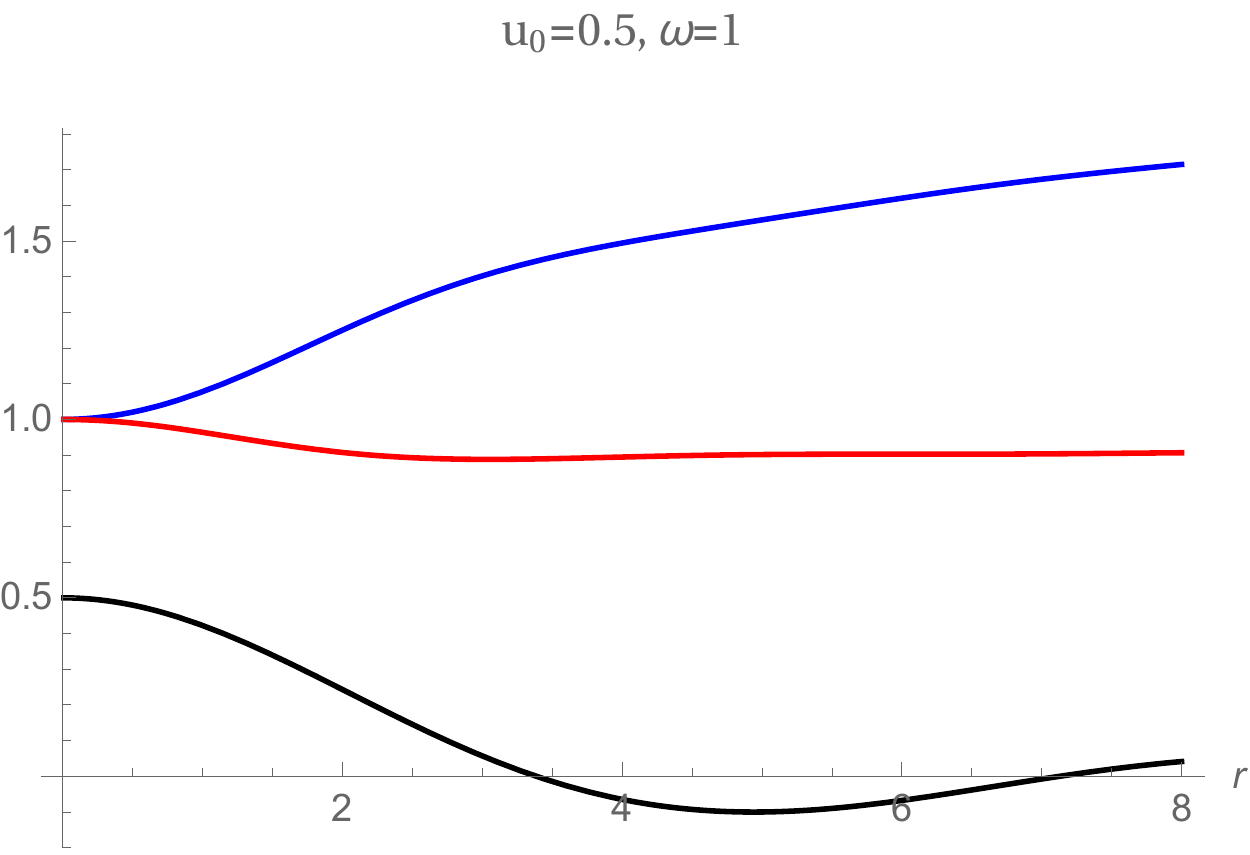}
\quad\includegraphics[scale=0.55]{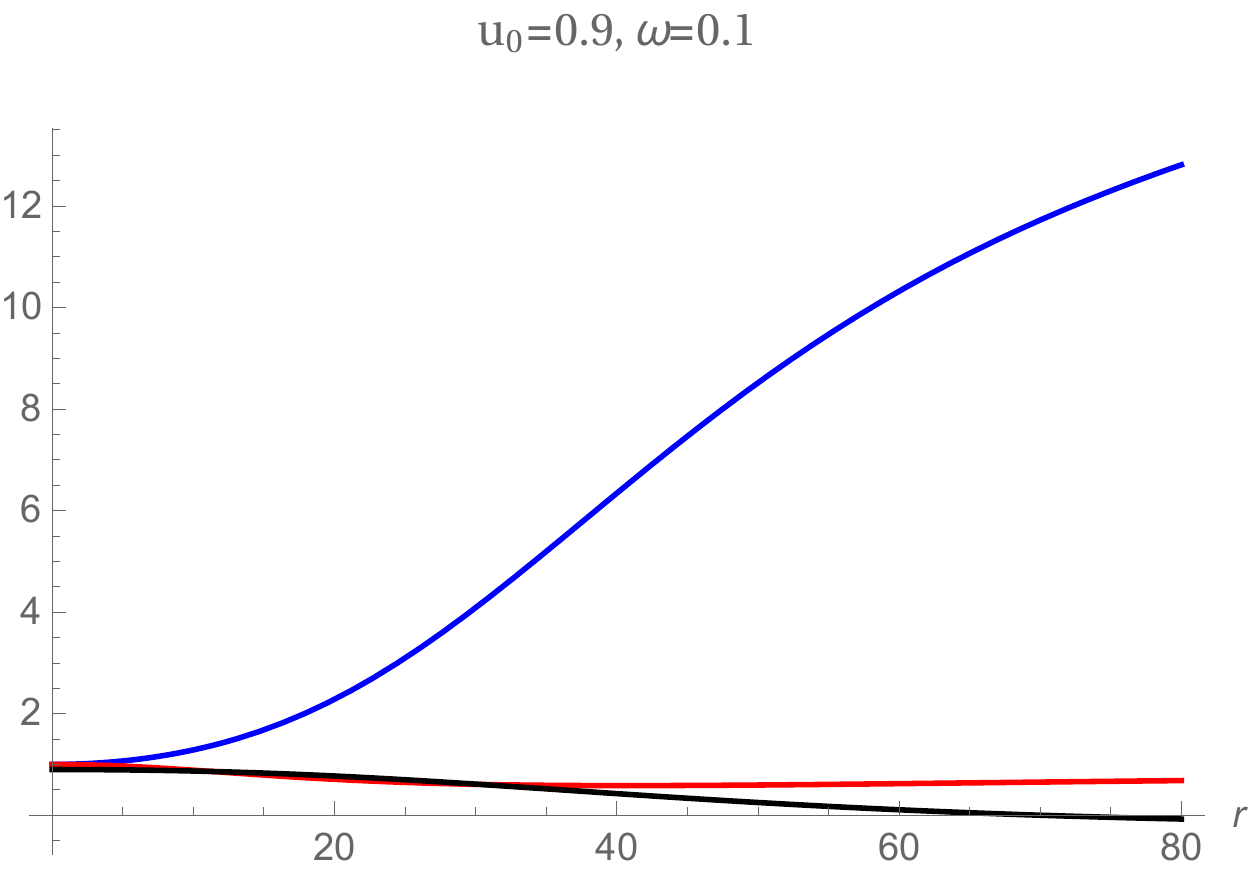}\quad\includegraphics[scale=0.55]{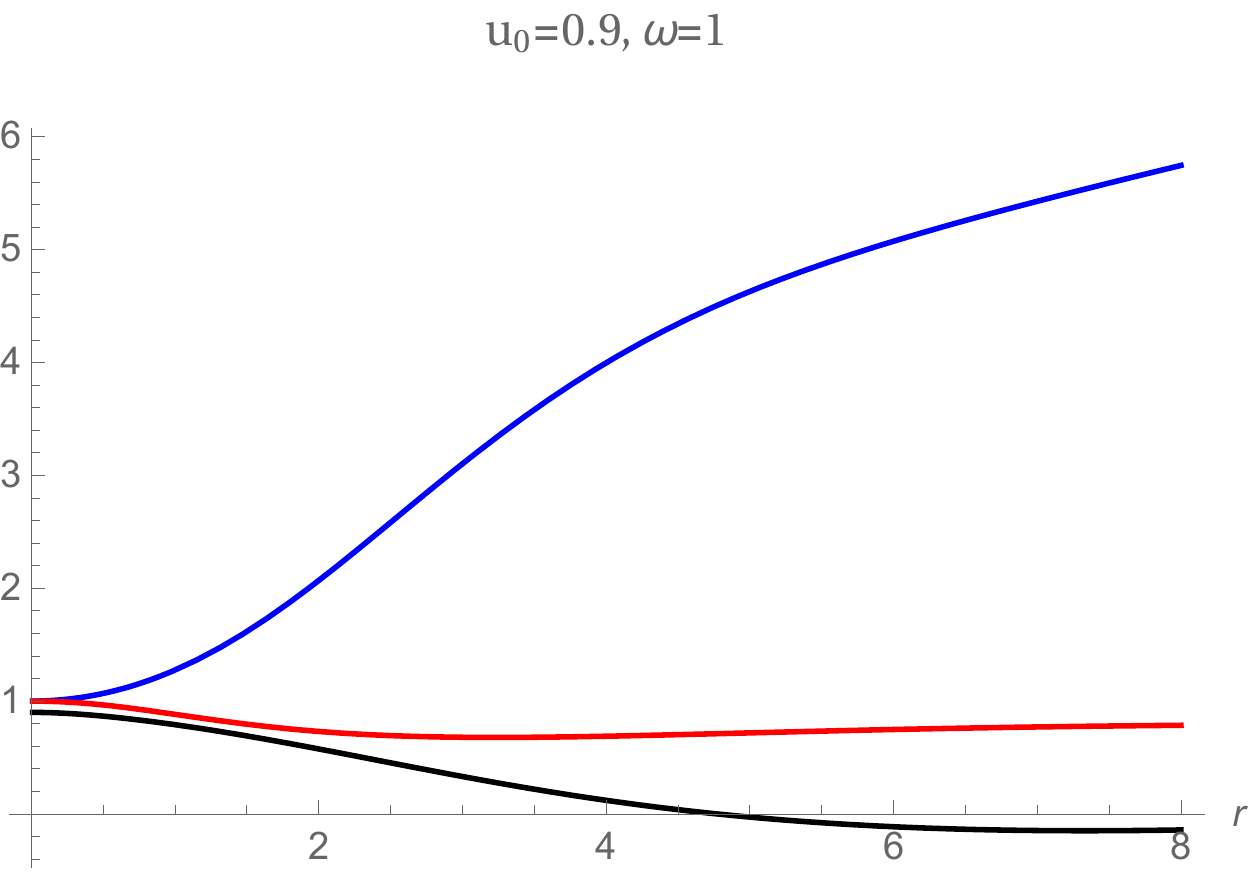}\caption{Metric functions as well as the Skyrme field for different values of $u_0$ and frequency for the Branch 1. The cavity wall is located at the first zero of the Skyrmion profile. We have set the Skyrme coupling $\lambda=1$.}\label{funcionesanalitica}
\end{figure}

\begin{figure}[h]
\centering
   \includegraphics[width=0.55\textwidth]{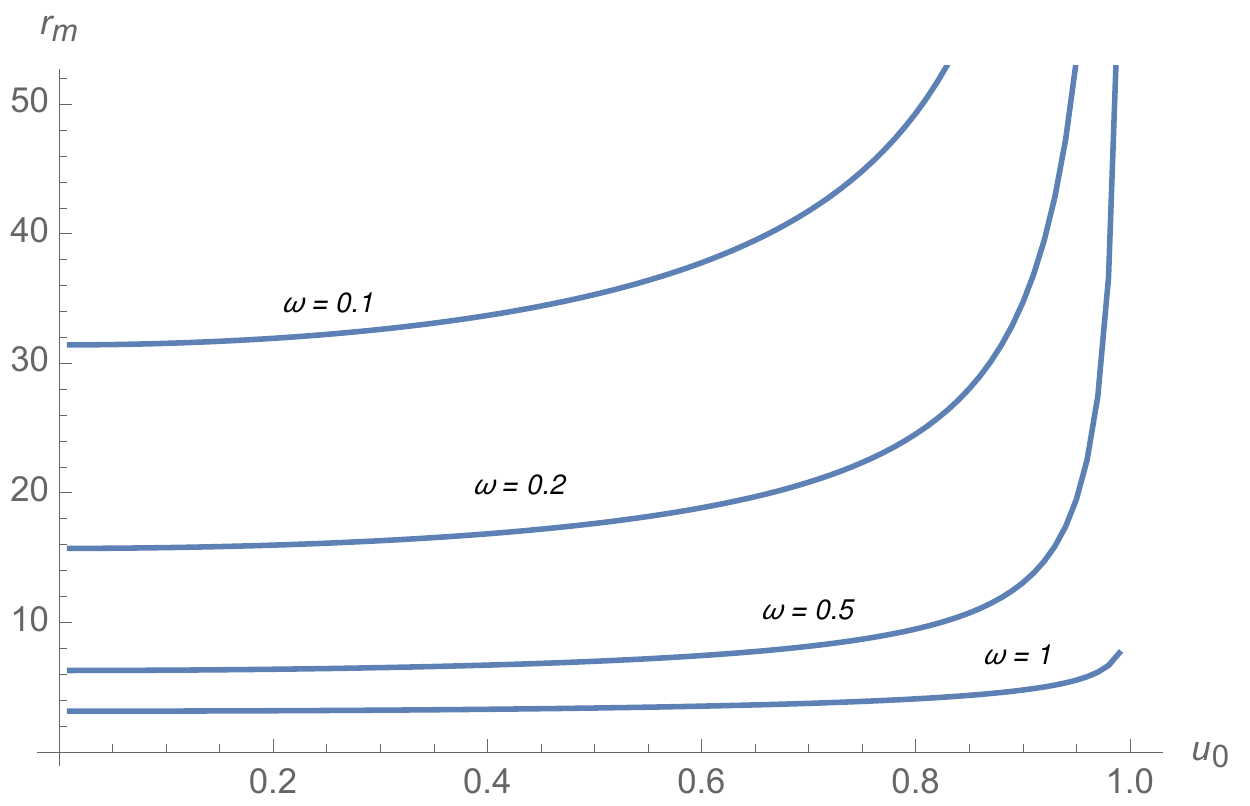}
  \caption{This figure depicts the radius of the mirror $r=r_m$ as an increasing function of the value of the Skyrme field at the origin for different values of $\omega$ (with $\lambda=1$) for the Branch 1.} \label{rmdewyu0}
\end{figure}

\begin{figure}[H]
  \centering
   \includegraphics[width=0.45\textwidth]{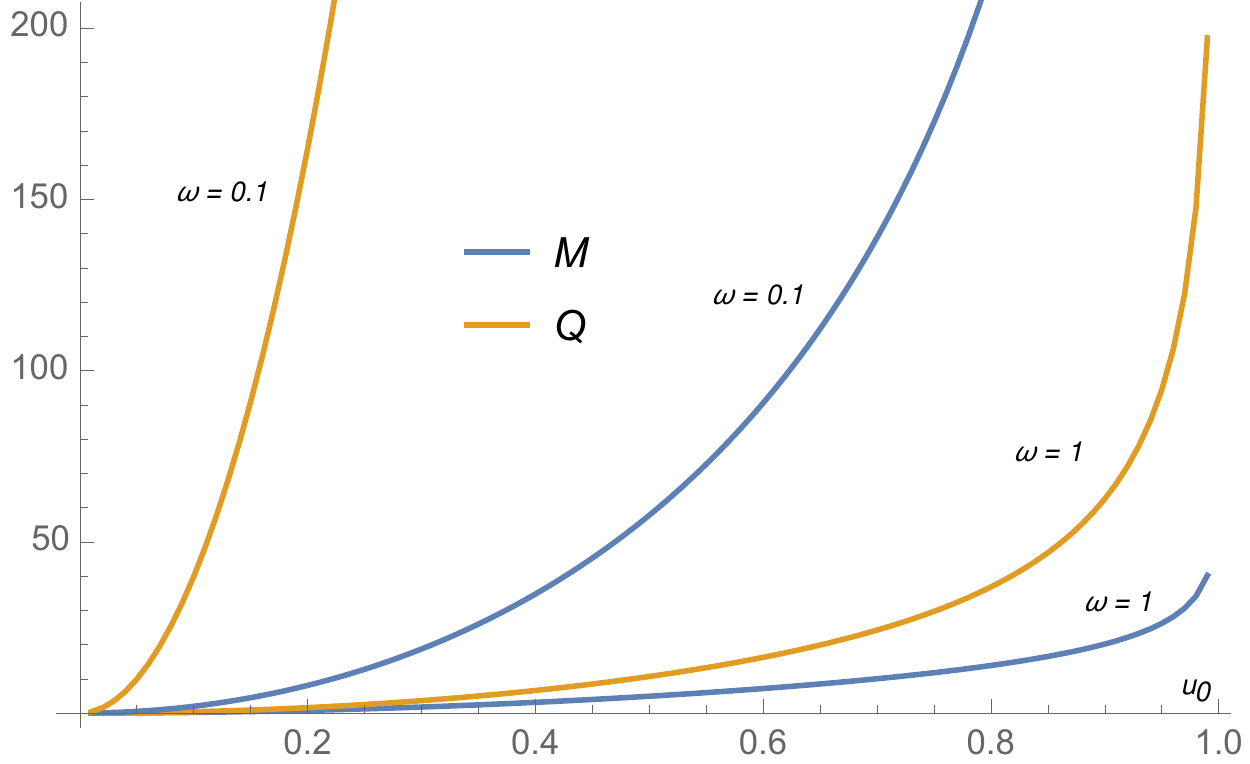}\qquad \includegraphics[width=0.45\textwidth]{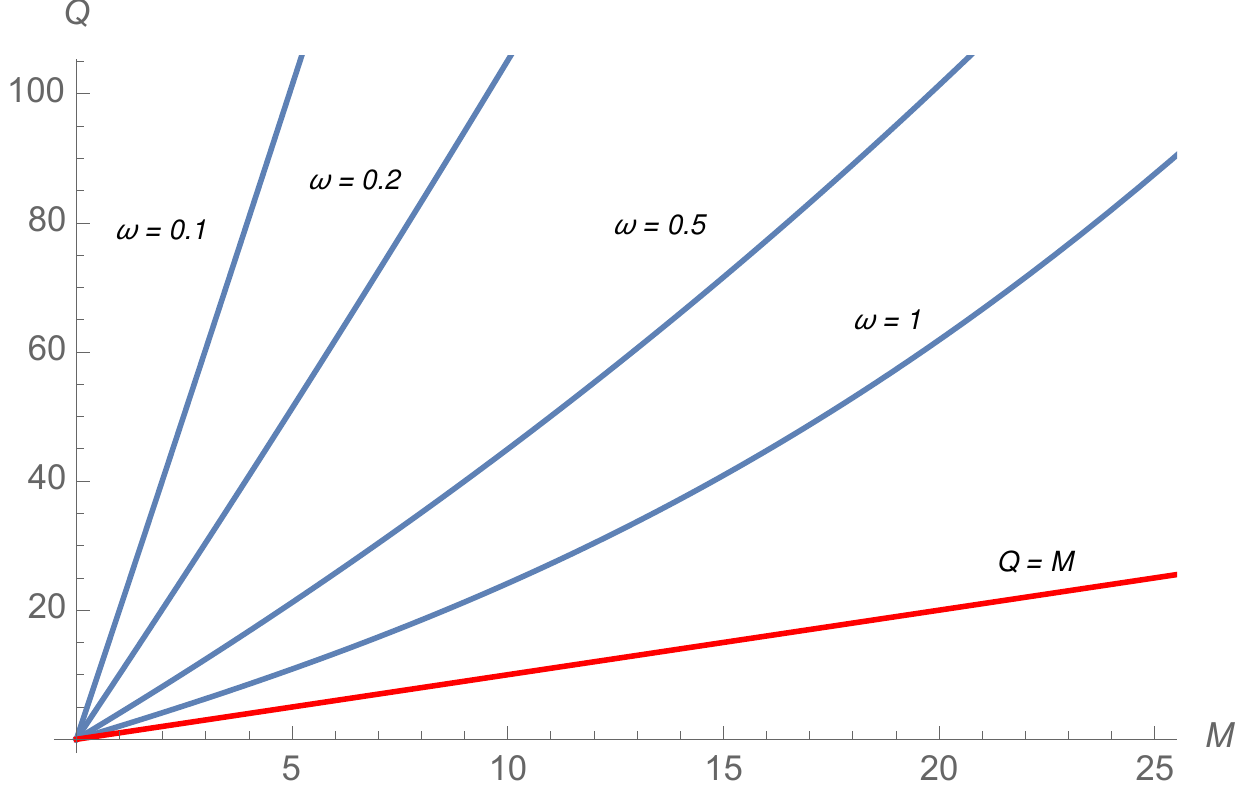}
  \caption{\textbf{Left panel}: The energy and the $U(1)$ charge of the solitons for two values of the frequency and a variety of values of the field at the origin for Branch 1. As expected, the charges diverge as $u_0\rightarrow 1$. \textbf{Right panel}: shows the behavior of $Q$ vs $M$ for the solitons in Branch 1. For all the cases $Q>M$ as can be seen by comparing with the red curve ($M=Q$) that has been included only as a reference. We have considered 100 values of $u_0$ in each curve, and its value increases in the range $0<u_0<1$ as one departs from the origin.}
  \label{QMdeu0}
\end{figure}

For the first branch, the free parameters are $u_0$ and $\omega$. Figure \ref{funcionesanalitica} shows the functions integrated from the system for four different combinations of the frequency and the strength 
of the Skyrme field at the origin. The mirror is located at the first zero of the black curve which represents the field $u(r)=\arcsin\alpha(r)$. The dependence of the radius of the mirror $r=r_m$ as a 
function of $u_0$ and the frequency is depicted in Figure \ref{rmdewyu0}. The radius of the mirror is an increasing function of the value of the field at the origin and increases with $\omega^{-1}$. The later 
is expected from the asymptotic behavior since the periodicity of the zeros of the field $\alpha(r)\sim \cos(\omega r)/r$ is locked in terms of the time periodicity of the phase $\Theta=\omega t$. 
Figure \ref{rmdewyu0} also shows that one could locate the mirror at an arbitrarily large proper distance from the origin as $u_0$ approaches to $1$,  notwithstanding as seen in the left panel of Figure \ref{QMdeu0} 
as $u_0\rightarrow 1$ the energy and the charge diverge, as expected from the asymptotic analysis, therefore only mirrors located at a finite proper distance from the origin are compatible with having finite energy 
and charge. As can be seen from the right panel of Figure \ref{QMdeu0}, for all the solitons obtained here the $U(1)$ charge is larger than the energy of the configuration.

\newpage

\subsection{Branch Non-Analytic in $\lambda$}
\begin{figure}[H]
\centering
\includegraphics[scale=0.7]{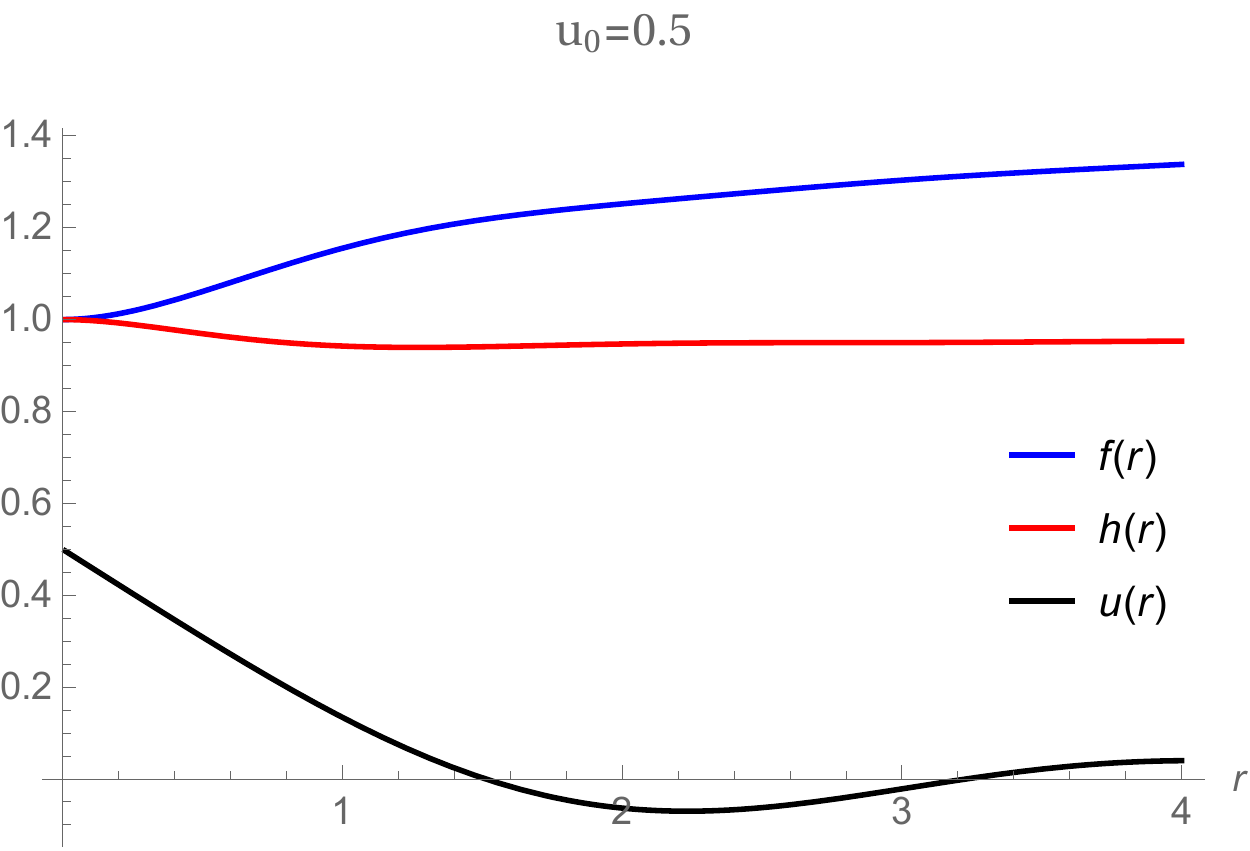}
\\
\includegraphics[scale=0.7]{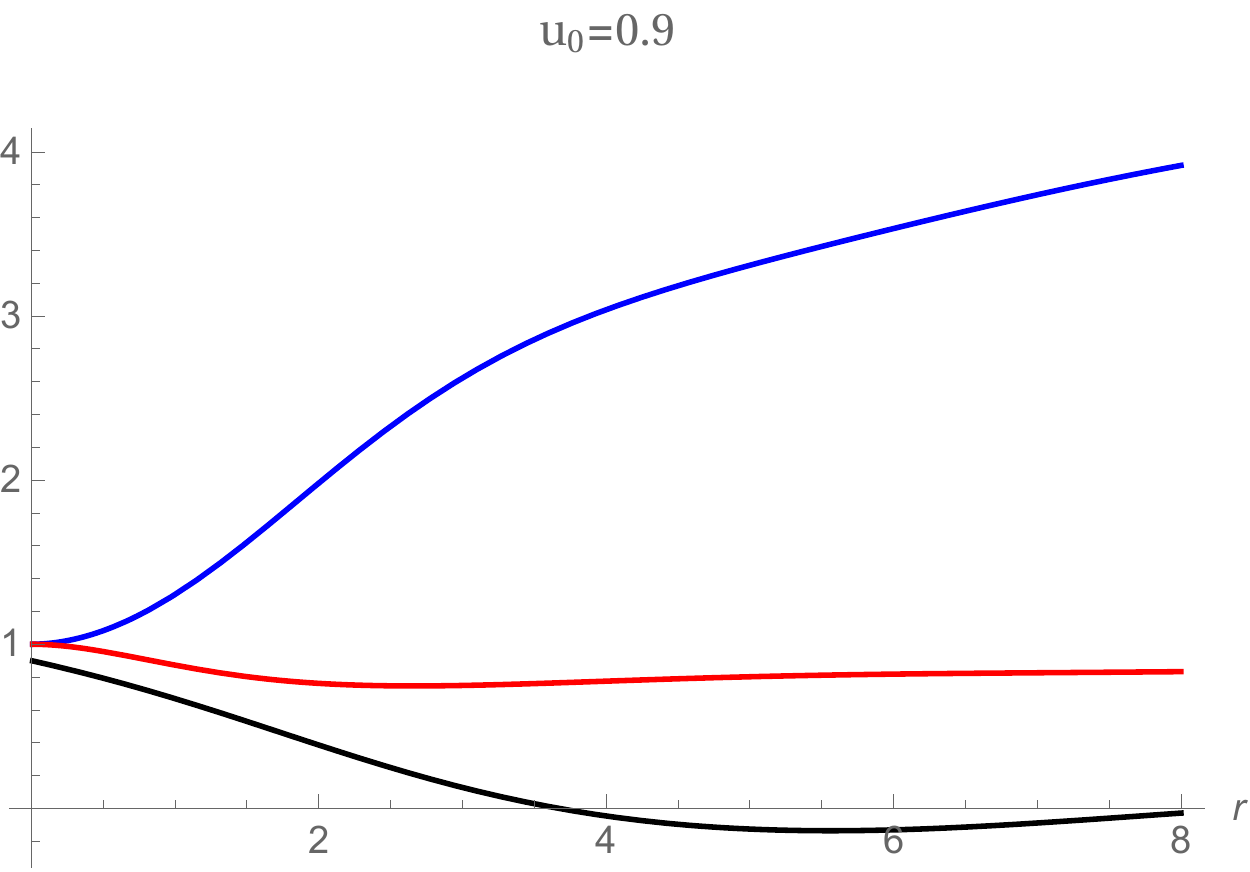}
\caption{Metric functions as well as the Skyrme field of the second branch, for different values of $u_0$. The lapse function has been set to $1$ at the origin which locks the frequency of the phase of the Skyrme field.}%
\label{curvasnon}
\end{figure}

As mentioned above, the second branch is non-analytic in the Skyrme coupling and it is characterized by a unique integration constant $u_0$ after one sets the time coordinate to coincide with the proper time of a geodesic observer located at the origin. Figure \ref{curvasnon} depicts the behavior of the metric functions as well as the Skyrme profile for different values of the latter at the origin. 

\begin{figure}[H]
  \centering
   \includegraphics[scale=0.55]{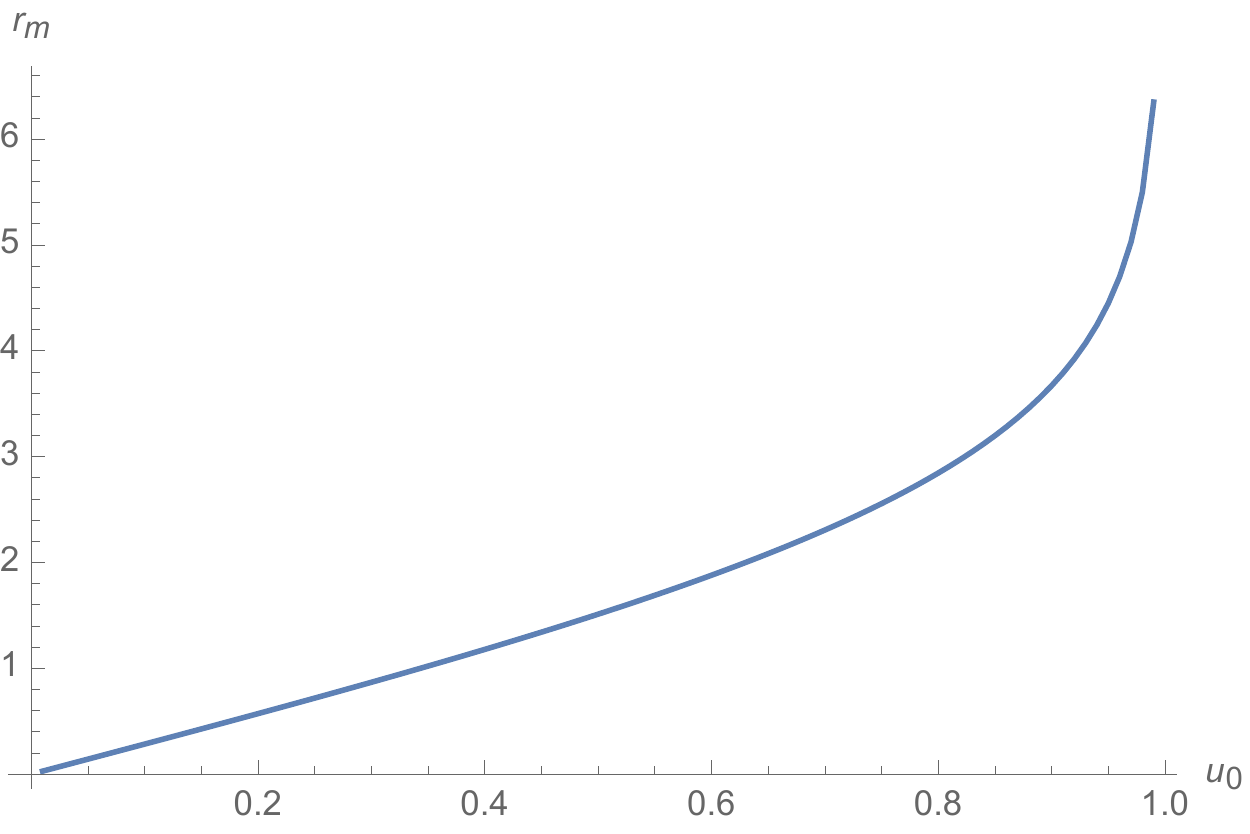}\quad\includegraphics[scale=0.55]{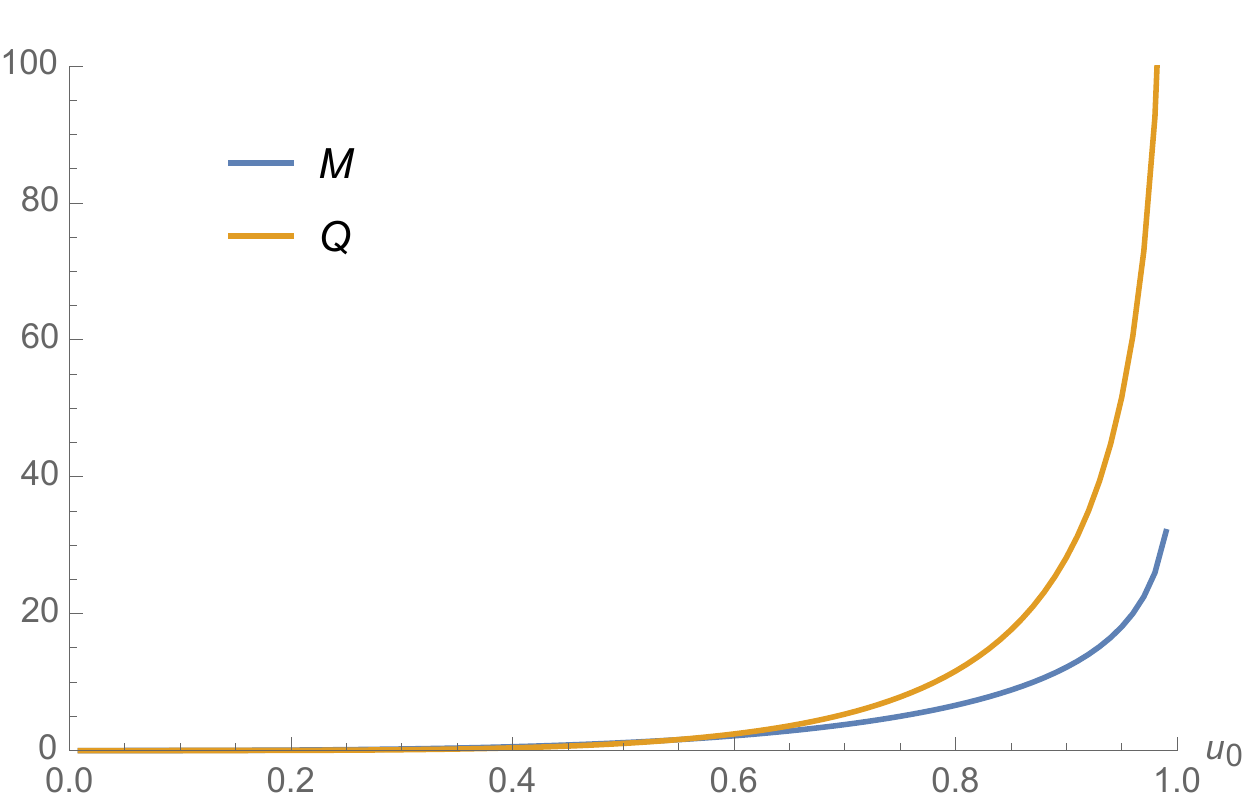}\quad \includegraphics[scale=0.55]{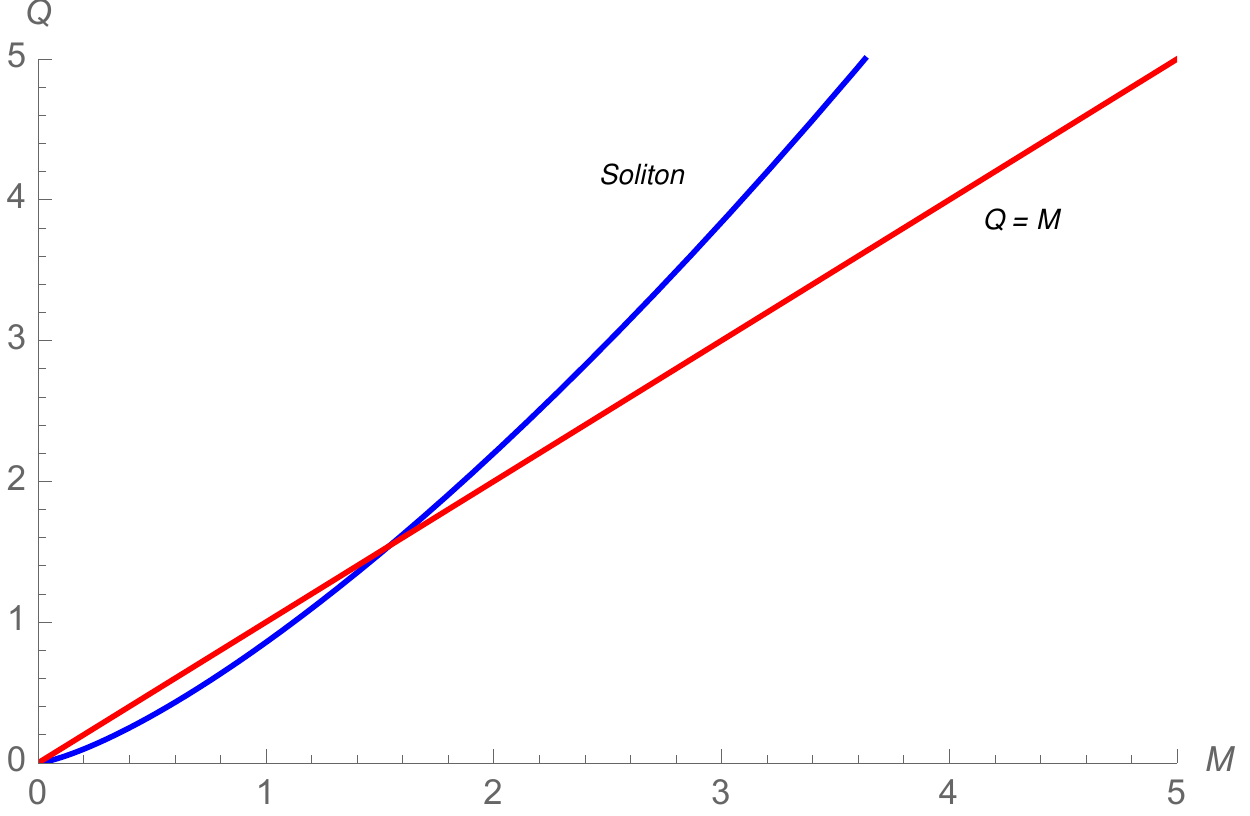}
  \caption{\textbf{Upper left panel:} The radius of the mirror is an increasing function of the value of $u_0$ for the non-analytic branch and it diverges as $u_0\rightarrow 1$. \textbf{Upper right panel:} $M$ and $Q$ vs $u_0$. Both diverge as the $r_m\rightarrow\infty$. \textbf{Lower panel:} $Q$ vs $M$. There is a critical value above which the charge is greater than the mass ($Q=M$ included only for reference and $\lambda=1$).}
  \label{rmdeu0non}
\end{figure}

Upper left pannel of Figure \ref{rmdeu0non} shows the behavior of the radius of the mirror as a function of the amplitude of the Skyrme field at the origin. Again, the radius diverges at $u_0$ approaches $1$ but as shown in the upper right Figure \ref{rmdeu0non} the mass and charge would diverge in that case. For small values of the mass and charge the curves seem to overlap. Lower panel of Figure \ref{rmdeu0non} shows that indeed there is a critical value for $u_0$ above which the charge surpasses the value of the mass, while below this critical value the mass is larger than the charge. In that figure we have included the curve $Q=M$ only for reference.

\section{Conclusions and further comments}
In this paper we have constructed new solutions of the Einstein-Skyrme model for $SU(2)$ group. We make use of the Generalized Hedgehog Ansatz turning on the fields along the $S^2\subset S^3$ submanifold. In the absence of the Skyrme term the system effectively reduces to a Non-linear Sigma Model on $S^2$. A cavity has been included which is located at the first zero of the Skyrme profile, and we studied the behavior of the mass and $U(1)$ charge as a function of the location of the boundary. The conserved charges for the different cases can be compared since all these configurations share the region located at the origin which allows to define a common normalization for the globally timelike Killing vector $\partial_t$. The regularity of the solutions at the origin imply the existence of two branches of solutions, and while the first branch exists for any value of the Skyrme coupling, the existence of the second branch is intrinsic to the presence of the term introduced by Skyrme to stabilize the solitons. After normalizing the time coordinate in order it to coincide with the proper time of a geodesic observer located at the origin one is left with solutions parameterized by two constants $(u_0,\omega)$ in the first branch and by a single constant $u_0$ in the second branch. In the former case we observe that the charge is always greater than the mass, while in the latter the charge is larger than the mass only above a critical value of the mass which induces a lower critical value for the amplitude of the Skyrme field at the origin.  
\\
\\
One might be tempted to construct black holes in a cavity with non-vanishing Skyrme profile in this system. Assuming the existence of a regular horizon located at $r=r_+$, as well as assuming analyticity for $u(r)$ at the horizon, one can show that the field equations have two branches. In the first branch $u(r)=0$, which implies $U$ equals the identity of $SU(2)$, and the expansions of functions $f$ and $g$ reconstruct the Schwarzschild solutions. The second branch leads to a near horizon expansion of the form
\begin{subequations}\label{bhsys}
\renewcommand{\theequation}{\theparentequation.\arabic{equation}}
\begin{align}
u(r)&=u_1 (r-r_+)+\mathcal{O}((r-r_+)^2)\ ,\\
g(r)&=r_+^{-1}(r-r_+)+\mathcal{O}((r-r_+)^2)\ ,\\
f(r)&=-r_+\omega^2(r-r_+)+\mathcal{O}((r-r_+)^2)\ .
\end{align}
\end{subequations}
which is not consistent with the structure of an event horizon. This shows that, within the ansatz here considered, there are no non-trivial, black hole solutions. Therefore, the boson stars constructed in this work cannot decay into a hairy black hole with the same symmetries, because such black hole does not exist\footnote{Boson stars with a local $SU(2)$ symmetry in the context of Einstein-Yang-Mills are constructed in \cite{Brihaye:2004nd}.}.
 It is interesting to note that the author of reference \cite{Verbin:2007fa} constructed boson star solutions in the Non-linear Sigma Model case by adding a designed self-interacting potential to (\ref{effaction}) without further constraint\footnote{This Lagrangian can also be seen as a member of the bi-scalar extension of Horndeski theories \cite{Horndeski:1974wa}. For recent constructions of boson stars in such setup see e.g. \cite{Brihaye:2015veu} and \cite{Brihaye:2016lin}.}. The presence of the self-interaction allows to construct configurations of finite mass even when the boundary of the cavity is located at an infinite proper distance from the origin. It would be interesting to include the self-interaction also for a finite cavity.

\section*{Acknowledgements}

M.L. and A.V. appreciates the support of CONICYT Fellowship 21141229 and
21151067, respectively. J. O. thanks Andrés Anabalón, Adolfo Cisterna and Carlos Herdeiro for enlightening discussions. A. V. thanks Gustavo Álvarez for useful discussions. 
This work was also funded by CONICYT Grant DPI20140053 (J.O. and A.G.) and FONDECYT Grant 1150246 (A.G.).

\end{document}